\documentclass[%
preprint,
superscriptaddress,
 amsmath,amssymb,
 aps,
]{revtex4-2}

\usepackage{graphicx}% Include figure files
\usepackage{dcolumn}% Align table columns on decimal point
\usepackage{bm}% bold math
\usepackage{hyperref}% add hypertext capabilities
\hypersetup{
    colorlinks=true,
    urlcolor= blue,
    citecolor=blue,
linkcolor= blue}

\begin{document}

\preprint{APS/123-QED}

\title{Mapping the phase diagram of the quantum anomalous Hall and topological Hall effects in a dual-gated magnetic topological insulator heterostructure}% Force line breaks with \\
\author{Run Xiao$^{\dagger}$}
\author{Di Xiao$^{\dagger}$}
\author{Jue Jiang$^{\dagger}$}
\author{Jae-Ho Shin}
\author{Fei Wang}
\author{Yi-Fan Zhao}
\author{Ruo-Xi Zhang}
\affiliation{Department of Physics, The Pennsylvania State University, University Park PA 16802}
\author{Anthony Richardella}
\affiliation{Department of Physics, The Pennsylvania State University, University Park PA 16802}
\affiliation{Materials Research Institute, The Pennsylvania State University, University Park PA 16802}
\author{Ke Wang}
\affiliation{Materials Research Institute, The Pennsylvania State University, University Park PA 16802}
\author{Morteza Kayyalha}
\affiliation{Department of Electrical Engineering, The Pennsylvania State University, University Park PA 16802}
\author{Moses H. W. Chan}
\author{Chao-Xing Liu}
\author{Cui-Zu Chang}
\thanks{Corresponding authors: nsamarth@psu.edu, cxc955@psu.edu}
\author{Nitin Samarth}
\thanks{Corresponding authors: nsamarth@psu.edu, cxc955@psu.edu}
\affiliation{Department of Physics, The Pennsylvania State University, University Park PA 16802}

%\def\thefootnote{$\dagger$}\footnotetext{These authors contributed equally to this work}\def\thefootnote{\arabic{footnote}}
%text text text\footnote{normal footnote}

\date{\today}% It is always \today, today,
             %  but any date may be explicitly specified

\begin{abstract}
We use magnetotransport in dual-gated magnetic topological insulator heterostructures to map out a phase diagram of the topological Hall and quantum anomalous Hall effects as a function of the chemical potential (primarily determined by the back gate voltage) and the asymmetric potential (primarily determined by the top gate voltage). A theoretical model that includes both surface states and valence band quantum well states allows the evaluation of the variation of the Dzyaloshinskii-Moriya interaction and carrier density with gate voltages. The qualitative agreement between experiment and theory provides strong evidence for the existence of a topological Hall effect in the system studied, opening up a new route for understanding and manipulating chiral magnetic spin textures in real space.
%\begin{description}
%\item[DOI]
%some information here.
%\end{description}
\end{abstract}

\maketitle
%\tableofcontents

%\section{\label{sec:level1}Introduction}
In recent years, condensed matter physics has seen a growing interest in studying the interplay between topology in momentum space and topology in real space. The former often manifests in nontrivial band structures in momentum space arising from the combined effects of some fundamental symmetry and strong spin-orbit coupling, while the latter (also a product of spin-orbit coupling) is associated with chiral magnetic spin textures in real space \cite{hasan2010colloquium,qi2011topological,nagaosa2013topological}. The quantum anomalous Hall (QAH) effect \cite{haldane1988model, liu2008quantum, qi2008topological, yu2010quantized, chang2013experimental,chang2015high,checkelsky2014trajectory,kou2014scale,mogi2015magnetic}, induced by a non-trivial Berry curvature in a topological system with broken time-reversal symmetry, provides convincing evidence of topology in momentum space. It is characterized by a quantized Hall resistance and a vanishing longitudinal resistance at zero magnetic field and has been realized in magnetically doped topological insulators (TIs) \cite{chang2013experimental,chang2015high,checkelsky2014trajectory,kou2014scale,mogi2015magnetic}. The topological Hall effect (THE), induced by the interaction of itinerant charge carriers with chiral spin textures such as magnetic skyrmions or chiral domain walls, is regarded as a signature of topology in real space \cite{nagaosa2013topological}. The THE manifests as an excess Hall voltage superimposed on the usual hysteretic anomalous Hall voltage that arises in magnetic conductors. Such a signature has been observed and interpreted as evidence for the THE in many systems, including MnSi \cite{neubauer2013erratum,lee2009unusual}, MnGe \cite{kanazawa2011large}, FeGe \cite{huang2012extended}, SrIrO$_3$/SrRuO$_3$ interface \cite{matsuno2016interface,ohuchi2018electric}, magnetically doped TI heterostructures \cite{yasuda2016geometric,liu2017dimensional,doi:10.1021/Wang_nanolett2021}, and TI/BaFe$_{12}$O$_{19}$ heterostructures \cite{Li_nanolett2021}. Given this context, it is valuable to identify model systems wherein the QAH and THE can be systematically studied as a function of some easily tuned system parameters. We recently studied the THE in one such model system, TI sandwich heterostructures of (Cr$_{0.15}$(Bi,Sb)$_{1.85}$Te$_3$ - (Bi,Sb)$_2$Te$_3$ - Cr$_{0.15}$(Bi,Sb)$_{1.85}$Te$_3$, referred as CBST-BST-CBST below) \cite{jiang2020concurrence}. By using a bottom gate to tune the chemical potential, we showed how a single sample could be continuously tuned from the QAH effect regime to the THE regime. 

In this {\it Letter}, we further extend the tunability of this model system by adding a top gate and demonstrate how a dual gating scheme enables the mapping of a phase diagram of the concurrence of the QAH effect and the THE as a function of the chemical potential and the asymmetry in the potential between the top and bottom surfaces. In particular, we find that the THE is enhanced when top and bottom gate voltages have different signs and that it is quenched when these gate voltages have the same sign. We also demonstrate that the THE arises because the asymmetric potential induced by dual gates leads to a Dzyaloshinskii-Moriya (DM) interaction.

%\section{\label{sec:level2}Experimental}
We used a VEECO 620 molecular beam epitaxy system to grow 3 quintuple layer (QL) CBST - 5QL BST - 3QL CBST heterostructures on SrTiO$_3$ (111) substrates (MTI Corporation). The SrTiO$_3$ substrates were first soaked in deionized  water at $90^\circ$C for 1.5 hours and thermally annealed at $985^\circ$C  for 3 hours in a tube furnace with flowing oxygen gas. The heat-treated  SrTiO$_3$ substrates were then outgassed under vacuum at $630^\circ$C (thermocouple temperature) for 1 hour. After outgassing, the substrates were cooled down to $340^\circ$C (thermocouple temperature) for the heterostructure growth. High purity Cr (5 N), Bi (5 N), Sb (6 N), and Te (6 N) were evaporated from Knudsen effusion cells. The cell temperatures were precisely controlled to obtain the desired beam equivalent pressure (BEP) fluxes of each element. The BEP flux ratio of Te/(Bi+Sb) was higher than 10 to prevent Te deficiency. The BEP flux ratio of Sb/Bi was around 2 to tune the heterostructure's chemical potential close to the charge neutral point. The heterostructure growth rate was \~{} 0.25 QL/min, and the pressure of the MBE chamber was maintained at $2 \times 10^{-10}$ mbar during the growth.

After the growth, the heterostructures were capped with 10 nm Te \textit{in situ} at $20^\circ$C for protection during the fabrication process.  Heterostructures were then fabricated into Hall bar devices using photolithography and Ar$^+$ plasma dry etching. The top gate was defined by a 40 nm Al$_2$O$_3$ dielectric layer and 5nm Ti/50nm Au contacts deposited by atomic layer deposition and electron beam evaporation, respectively.

We carried out magnetotransport measurements in a commercial He3 fridge (Oxford Heliox) for temperatures higher than 0.4 K and in a commercial Leiden Cryogenics dilution refrigerator at $T$ = 60 mK.  Bottom and top gate voltages were applied using the SrTiO$_3$ substrate and the deposited Al$_2$O$_3$ layer as the dielectric layers. 

%\section{\label{sec:level3}Results and discussion}

\begin{figure*}
\includegraphics[width=0.9\textwidth]{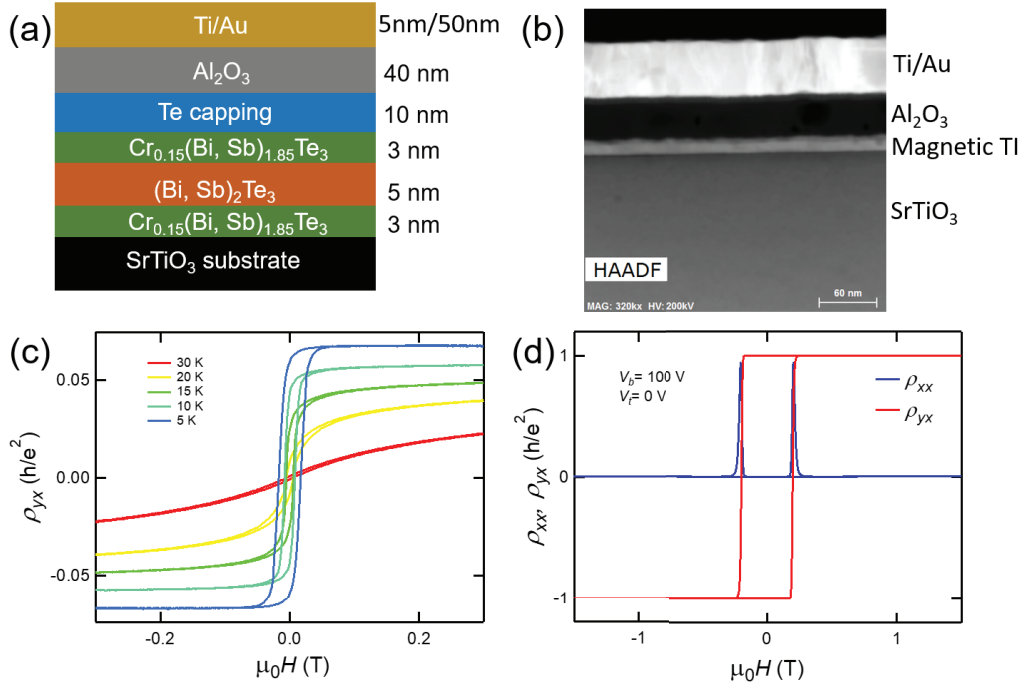}% Here is how to import EPS art
\caption{\label{FIG.1} The QAH effect in dual-gated magnetic TI heterostructures. (a) A schematic structure of the magnetic TI heterostructure. The compositions of the layers are nominal and based on past calibrations of sample growth. (b) Cross-sectional TEM image of a dual-gated device.  (c) Temperature variation of the magnetic field dependence of Hall resistance. (d) Demonstration of the QAH effect at $T$ = 60 mK with gate voltages $V_b=100$ V and $V_t=0$ V. 
}
\end{figure*}
Figures \ref{FIG.1}(a) and (b) show the detailed schematic structure and a cross-sectional transmission electron microscope (TEM) image, respectively, of one sample we  studied. The sandwich heterostructure geometry consists of two ultrathin (3 QL) magnetic TI layers separated by a non-magnetic TI layer and has been shown to improve the quality of the observed QAH effect at higher temperatures (up to 2 K)\cite{mogi2015magnetic}. More important for the purposes of this paper, since the magnetic TI layers are separated by a non-magnetic TI, the magnetic exchange interaction between top and bottom layers is weakened. This allows for a significant DM interaction, which is a requirement for the THE \cite{nagaosa2013topological}. 
Figure \ref{FIG.1}(c) shows the temperature dependence of anomalous Hall effect.  A hysteresis loop develops below a Curie temperature around 20 K. As the sample is cooled down to the base temperature at $T$ = 60 mK, a robust QAH effect develops depending on the values of the top and bottom gate voltages (V$_b$=100 V, V$_t$=0 V in Figs. \ref{FIG.1} (d) ).

\begin{figure*}
\includegraphics[width=0.9\textwidth]{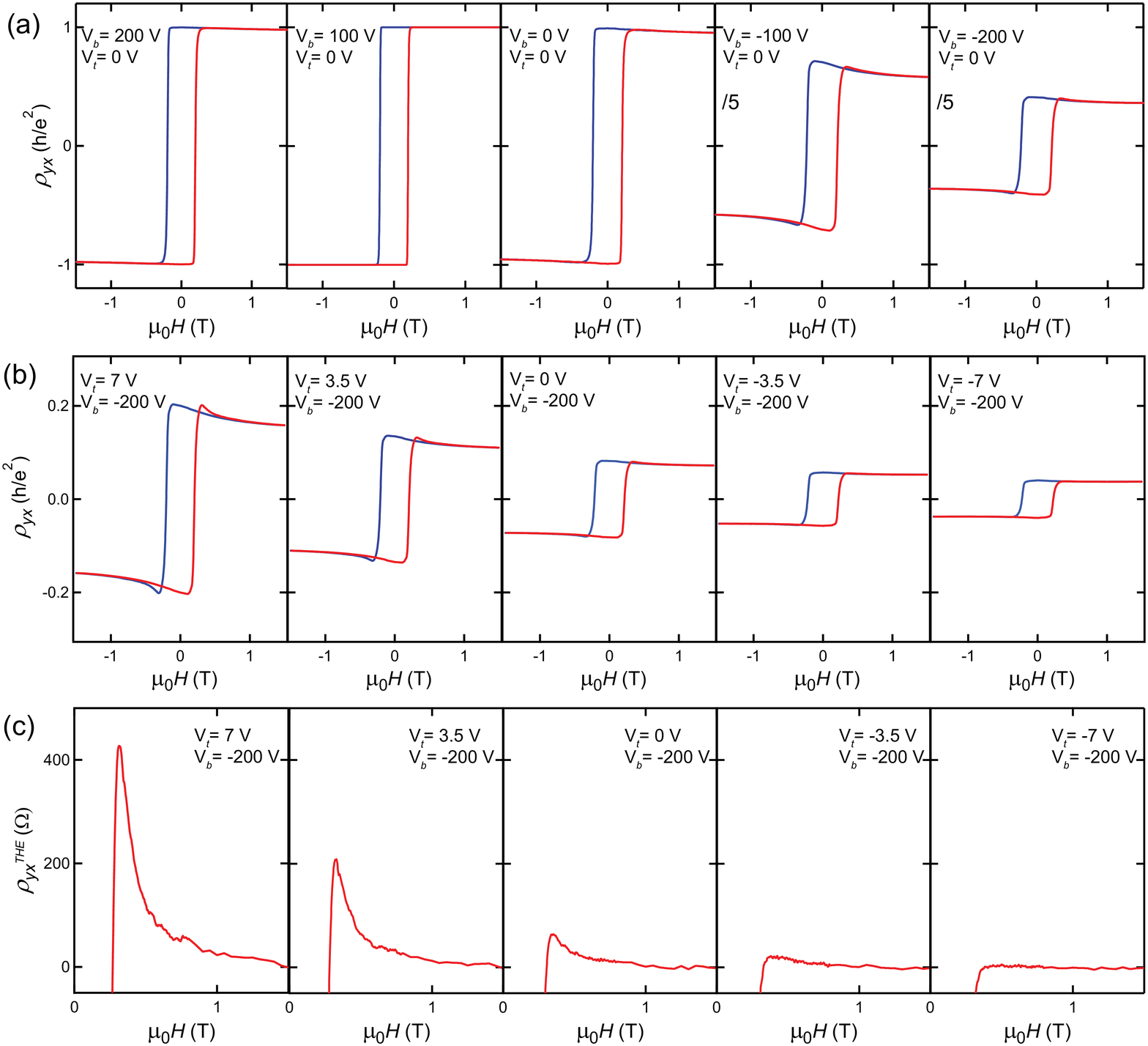}% Here is how to import EPS art
\caption{\label{FIG.2} The topological Hall effect observed in dual-gated magnetic TI heterostructures. (a) The deviation from the QAH regime using the bottom gate. (b) Top gate-voltage-driven THE. The bottom gate voltage V$_b$ is fixed at -200V. By applying a positive top gate voltage V$_t$, the THE becomes more pronounced. On the other hand, applying a negative V$_t$ can make the THE vanish. (c) Top gate voltage dependence of the magnitude of the THE $\rho_{yx}^{THE}$. The data are obtained by subtracting a constant high field Hall resistance. }
\end{figure*}
At the base temperature $T$ = 60 mK, we mapped out the top and bottom gate voltage dependence of the QAH effect and the THE. As shown in Fig. \ref{FIG.2}(a), if the bottom gate V$_b$ is changed while the top gate V$_t$ is fixed at 0 V, the heterostructure can be tuned away from the QAH regime, particularly at negative V$_b$. Away from the QAH regime, at magnetic fields larger than the coercive field, a careful examination of the Hall resistance shows that it is slightly different during the upward and downward magnetic field sweeps.  The excess Hall resistance after crossing the magnetic reversal transition is a  signature of the THE. Therefore, by tuning V$_b$, we observe a crossover from the QAH effect to the THE. This is due to the formation of chiral domain walls in the presence of a strong DM interaction \cite{jiang2020concurrence}.  

The signature of the THE becomes more obvious if we apply a positive V$_t$ while keeping V$_b$ fixed at a negative value (Fig. \ref{FIG.2}(b)). If we fix V$_b$ and change V$_t$, the THE behaves differently: unlike the case of $V_b$-modulated THE, it becomes more pronounced at positive V$_t$ but vanishes under negative V$_t$. In Fig. \ref{FIG.2}(c), we plot the gate voltage dependence of the magnitude of the THE, obtained by subtracting the Hall voltage measured during the upward and downward sweeps of the magnetic field ($\rho_{yx}^{THE}$). This allows us to observe the variation of the QAH effect and the THE as a function of the chemical potential. The maximum of $\rho_{yx}^{THE}$ occurs when V$_t$ and V$_b$ have different signs. In contrast, when they have the same sign, $\rho_{yx}^{THE}$ decreases and vanishes.

\begin{figure*}
\includegraphics[width=0.9\textwidth]{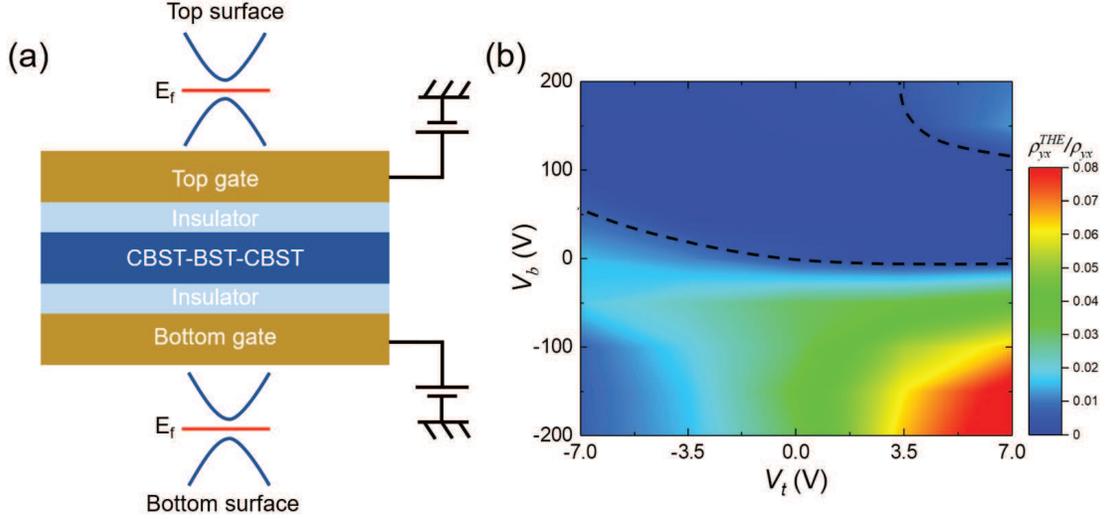}% Here is how to import EPS art
\caption{\label{FIG.3} The gate voltage dependence of the THE. (a)  Capacitance model of the gating effect. The top gate mainly tunes the chemical potential of the top surface while the bottom gate mainly tunes the chemical potential of the bottom surface. Dual gates can induce an asymmetric potential between top and bottom surfaces. (b) The THE phase diagram. THE is absent in the QAH effect regime (dark blue area). Away from the QAH regime, when the top and back gate voltages have different signs, the THE becomes more obvious (red area). When dual gates have the same sign, the THE is quenched (lower left corner).}
\end{figure*}

In order to understand the behavior of the THE, we propose a simple capacitance model (Fig. \ref{FIG.3}(a)). The top and bottom gates act as capacitors that inject or repel electrons in the top and bottom surfaces, respectively. Therefore, the top gate chiefly tunes the chemical potential of the top surface, while the bottom gate principally affects the chemical potential of the bottom surface. As a result, an asymmetric potential between top and bottom surfaces is induced, leading to the breaking of inversion symmetry in the sandwich heterostructure. Due to the broken inversion symmetry, the overall DM interaction from two surface is non-zero, giving rise to the THE. 

As shown in Fig. \ref{FIG.3}(b), the sandwich heterostructure enters the QAH effect regime in the dark blue area. In this region of the phase diagram, conduction from surface and bulk carriers is minimized; thus, the sample shows a perfect QAH effect with no THE signal. Away from the QAH regime where surface carriers start to contribute, a finite DM interaction is induced, and the THE appears. Furthermore, by creating an asymmetric potential by V$_b$ and V$_t$, $\rho_{yx}^{THE}$ reaches maximum shown in the red area.

\begin{figure*}
\includegraphics[width=0.9\textwidth]{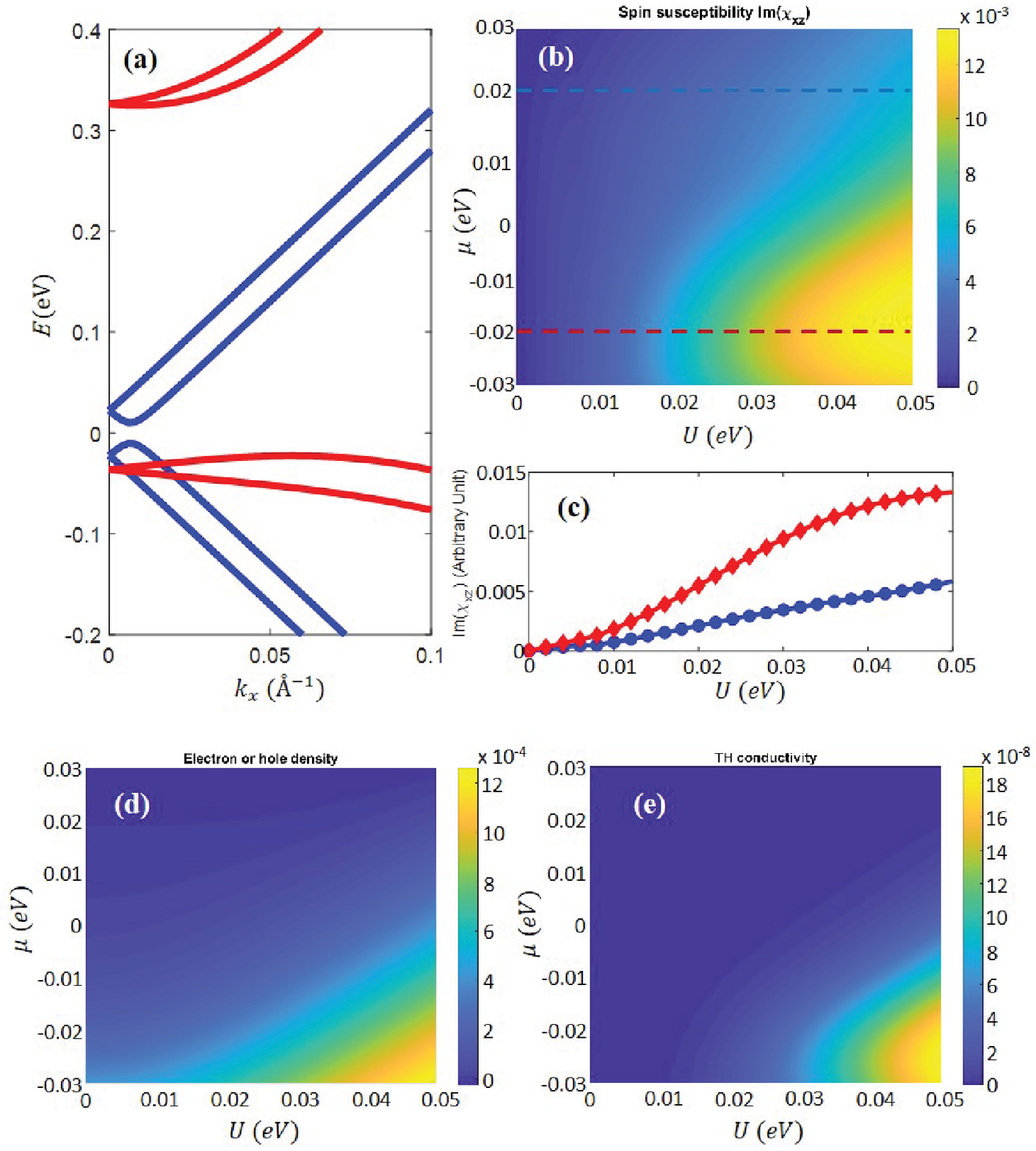}% Here is how to import EPS art
\caption{\label{FIG.4} (a) The energy dispersion of the surface states and bulk quantum well bands in a magnetic TI. Here $U=0.02eV$. (b) Spin susceptibility $\chi_{xz}$ as a function of chemical potential $\mu$ and asymmetric potential $U$. The blue and red dashed lines correspond to the blue and red plots in (c).  (c) Spin susceptibility $\chi_{xz}$ as a function of $U$ for the chemical potential $\mu=0.02eV$ (blue line) and $\mu=-0.02eV$ (red line). (d) Electron or hole carrier concentration $\rho$ as a function of chemical potential $\mu$ and asymmetric potential $U$. Here $\rho$ is in the unit of $\AA^-2$. The TH resistance is expected to proportional to $|\chi_{xz}|^2\rho$, whose dependence on $\mu$ and $U$ is shown in (e). Here we choose $q_x=0.005 \AA^{-1}$ and $q_y=0$.    }
\end{figure*}

%\section{\label{sec:theory}Theoretical model}
We now theoretically evaluate the DM interaction and the bulk carrier concentrations in magnetic TI heterostructures based on the model developed in Ref. \cite{jiang2020concurrence}. This model involves both the topological surface states and the bulk valence band quantum well states, both of which have been shown to be crucial in understanding the temperature and chemical potential dependence in the previous report \cite{jiang2020concurrence}. The topological surface states can be described by the effective Hamiltonian $H_{SS}=v_F(k_y\sigma_x-k_x\sigma_y)\tau_z+U\tau_z+m_0\tau_x+H_{SS,ex}$, where $\sigma$ is the Pauli matrices for the spin while $\tau$ stands for two surfaces. Here $v_F$ is the Fermi velocity, $U$ is the asymmetric potential and $m_0$ describes the hybridization between two surface states at the top and bottom surfaces. The exchange coupling between surface states and magnetic moments can be described by the Hamiltonian $H_{SS,ex}=-{\bf M}^t\cdot{\bf \sigma}\frac{1+\tau_z}{2}-{\bf M}^b\cdot{\bf \sigma}\frac{1-\tau_z}{2}$, where ${\bf M}^t$ and ${\bf M}^b$ label the magnetization at the top and bottom surfaces, respectively. The valence band quantum well states are described by the effective Hamiltonian $H_{QW}=\varepsilon_0(k)+N(k)\tau_z+A(k_y\sigma_x-k_x\sigma_y)\tau_x+U\tau_x+H_{QW,ex}$, where $\sigma$ still labels spin and $\tau$ stands for two orbitals instead. Here $\varepsilon_0(k)=C_0+C_1k^2$, $N(k)=N_0+N_1k^2$ and $C_0, C_1, N_0, N_1, A$ are material dependent parameters while $U$ is still for the asymmetric potential. The exchange coupling between quantum well states and magnetic moments is given by $H_{QW,ex}=-{\bf M}\cdot{\bf \sigma}$, where ${\bf M}$ is the magnetization. We assume the magnetization is uniform throughout the whole system and thus ${\bf M}^t={\bf M}^b={\bf M}$. As the strength of DM interaction is directly determined by the off-diagonal components of the spin susceptibility\cite{jiang2020concurrence}, particularly the components $\chi_{xz}$ and $\chi_{yz}$, we next discuss the behavior of $\chi_{xz}$ in our model ($\chi_{yz}$ can be directly related to $\chi_{xz}$ due to the rotation symmetry of our model). The zero-frequency spin susceptibility is given by $\chi_{\alpha\beta}(q)=\frac{1}{\beta}\sum_{i\omega_m,k}Tr[G_0(q+k,i\omega_m)\sigma_{\alpha}G_0(k,i\omega_m)\sigma_{\beta}]$ with $\beta=\frac{1}{k_BT}$ and $T$ is the temperature. Fig. \ref{FIG.4}(b) shows $\chi_{xz}$ as a function of asymmetric potential and chemical potential. The behavior of $\chi_{xz}$ can be understood from the energy dispersion in Fig. \ref{FIG.4}(a), which shows that the surface Dirac cones are close to the top of valence band quantum well states \cite{chang2015PRL,Wang2018NP}. The asymmetric potential can split the two spin states of the valence band and the surface states at the opposite surfaces. These splittings generally give rise to non-zero $\chi_{xz}$, as shown in Fig. \ref{FIG.4}(c) for two different chemical potentials. We also notice that when the Fermi energy lies between the two spin bands of the valence quantum well states, a larger $\chi_{xz}$ can be induced, as clearly shown in Fig. \ref{FIG.4}(c), in which the red line is for $\mu=-0.02eV$ that crosses the valence band top and the blue line is for $\mu=0.02eV$ that only crosses surface state. This is consistent with the experimental observation that the observed TH resistance is enhanced for negative back gate voltage $V_b$, which is expected to mainly tune the chemical potential to the valence bands. The front gate voltage $V_f$ is expected to mainly tune the asymmetric potential $U$ in our model and one can see from Fig. \ref{FIG.4}(c) that a large $U$ can give rise to a strong enhancement on $\chi_{xz}$. This is consistent with the observation in Fig. \ref{FIG.3}(b). 

We note that when the chemical potential crosses only the surface states, we still see a finite $\chi_{xz}$, but in experiments, the TH resistance almost vanishes. This is because the TH resistance can only come from the bulk carriers and it vanishes when the system is in the QAH regime with almost vanishing bulk carriers. Therefore, we also show the bulk carrier concentration $\rho$ in Fig. \ref{FIG.4}(d) and if we assume the TH resistance is proportional to both spin susceptibility $|\chi_{xz}|^2$ and bulk carrier concentration, the behavior of TH resistance is shown in Fig. \ref{FIG.4}(e).  

%\section{\label{sec:level4}Conclusions}

To summarize, we have studied the gate dependence of the THE in dual-gated magnetic TI sandwich heterostructures. We observed a crossover from the QAH effect to the THE by tuning the bottom gate, particularly under a negative bottom gate voltage. The magnitude of the THE increases with V$_b$ because this induces an asymmetric potential between the two opposite surfaces. By applying V$_t$ and V$_b$, this asymmetric potential and the magnitude of the THE can be enlarged or quenched by changing the relative sign of V$_t$ and V$_b$.  Since the THE in magnetic TI sandwich heterostructures provides evidence of chiral magnetic domain walls, the manipulation of the THE by dual gates provides a simple way to investigate and understand chiral magnetic spin textures in real space. We note that the good correspondence between our experimantal observations and theory rules against simpler explanations\cite{Filakowski_PhysRevX.10.011012} for the THE signals in our transport measurements. Our study will also motivate more studies on nontrivial quantum phenomena in magnetic TI multilayer heterostructures and facilitate the development of the proof-of-concept TI-based spintronic devices.

\begin{acknowledgments}
The synthesis of samples was supported by NSF 2DCC-MIP (DMR-1539916). The dilution electrical transport measurements were supported by DOE grant (DE-SC0019064), the Gordon and Betty Moore Foundation’s EPiQS Initiative (grant no. GBMF9063 to C.-Z. Chang), and and NSF (DMR-1707340). The electrical transport measurements above 400 mK were supported by the Institute for Quantum Matter under DOE EFRC grant DE-SC0019331.The theoretical calculations were also supported by DOE grant (DE-SC0019064).  

$^\dagger$R.X., D.X.and J.J contributed equally to this work.

\end{acknowledgments}

% The \nocite command causes all entries in a bibliography to be printed out
% whether or not they are actually referenced in the text. This is appropriate
% for the sample file to show the different styles of references, but authors
% most likely will not want to use it.
%\nocite{*}
%\bibliography{apssamp}% Produces the bibliography via BibTeX.

\begin{thebibliography}{26}%
\makeatletter
\providecommand \@ifxundefined [1]{%
 \@ifx{#1\undefined}
}%
\providecommand \@ifnum [1]{%
 \ifnum #1\expandafter \@firstoftwo
 \else \expandafter \@secondoftwo
 \fi
}%
\providecommand \@ifx [1]{%
 \ifx #1\expandafter \@firstoftwo
 \else \expandafter \@secondoftwo
 \fi
}%
\providecommand \natexlab [1]{#1}%
\providecommand \enquote  [1]{``#1''}%
\providecommand \bibnamefont  [1]{#1}%
\providecommand \bibfnamefont [1]{#1}%
\providecommand \citenamefont [1]{#1}%
\providecommand \href@noop [0]{\@secondoftwo}%
\providecommand \href [0]{\begingroup \@sanitize@url \@href}%
\providecommand \@href[1]{\@@startlink{#1}\@@href}%
\providecommand \@@href[1]{\endgroup#1\@@endlink}%
\providecommand \@sanitize@url [0]{\catcode `\\12\catcode `\$12\catcode
  `\&12\catcode `\#12\catcode `\^12\catcode `\_12\catcode `\%12\relax}%
\providecommand \@@startlink[1]{}%
\providecommand \@@endlink[0]{}%
\providecommand \url  [0]{\begingroup\@sanitize@url \@url }%
\providecommand \@url [1]{\endgroup\@href {#1}{\urlprefix }}%
\providecommand \urlprefix  [0]{URL }%
\providecommand \Eprint [0]{\href }%
\providecommand \doibase [0]{https://doi.org/}%
\providecommand \selectlanguage [0]{\@gobble}%
\providecommand \bibinfo  [0]{\@secondoftwo}%
\providecommand \bibfield  [0]{\@secondoftwo}%
\providecommand \translation [1]{[#1]}%
\providecommand \BibitemOpen [0]{}%
\providecommand \bibitemStop [0]{}%
\providecommand \bibitemNoStop [0]{.\EOS\space}%
\providecommand \EOS [0]{\spacefactor3000\relax}%
\providecommand \BibitemShut  [1]{\csname bibitem#1\endcsname}%
\let\auto@bib@innerbib\@empty
%</preamble>
\bibitem [{\citenamefont {Hasan}\ and\ \citenamefont
  {Kane}(2010)}]{hasan2010colloquium}%
  \BibitemOpen
  \bibfield  {author} {\bibinfo {author} {\bibfnamefont {M.~Z.}\ \bibnamefont
  {Hasan}}\ and\ \bibinfo {author} {\bibfnamefont {C.~L.}\ \bibnamefont
  {Kane}},\ }\bibfield  {title} {\bibinfo {title} {Colloquium: Topological
  insulators},\ }\href {https://doi.org/10.1103/RevModPhys.82.3045} {\bibfield
  {journal} {\bibinfo  {journal} {Rev. Mod. Phys.}\ }\textbf {\bibinfo {volume}
  {82}},\ \bibinfo {pages} {3045} (\bibinfo {year} {2010})}\BibitemShut
  {NoStop}%
\bibitem [{\citenamefont {Qi}\ and\ \citenamefont
  {Zhang}(2011)}]{qi2011topological}%
  \BibitemOpen
  \bibfield  {author} {\bibinfo {author} {\bibfnamefont {X.-L.}\ \bibnamefont
  {Qi}}\ and\ \bibinfo {author} {\bibfnamefont {S.-C.}\ \bibnamefont {Zhang}},\
  }\bibfield  {title} {\bibinfo {title} {Topological insulators and
  superconductors},\ }\href
  {https://journals.aps.org/rmp/abstract/10.1103/RevModPhys.83.1057} {\bibfield
   {journal} {\bibinfo  {journal} {Rev. Mod. Phys.}\ }\textbf {\bibinfo
  {volume} {83}},\ \bibinfo {pages} {1057} (\bibinfo {year}
  {2011})}\BibitemShut {NoStop}%
\bibitem [{\citenamefont {Nagaosa}\ and\ \citenamefont
  {Tokura}(2013)}]{nagaosa2013topological}%
  \BibitemOpen
  \bibfield  {author} {\bibinfo {author} {\bibfnamefont {N.}~\bibnamefont
  {Nagaosa}}\ and\ \bibinfo {author} {\bibfnamefont {Y.}~\bibnamefont
  {Tokura}},\ }\bibfield  {title} {\bibinfo {title} {Topological properties and
  dynamics of magnetic skyrmions},\ }\href
  {https://www.nature.com/articles/nnano.2013.243} {\bibfield  {journal}
  {\bibinfo  {journal} {Nat. Nanotechnol.}\ }\textbf {\bibinfo {volume} {8}},\
  \bibinfo {pages} {899} (\bibinfo {year} {2013})}\BibitemShut {NoStop}%
\bibitem [{\citenamefont {Haldane}(1988)}]{haldane1988model}%
  \BibitemOpen
  \bibfield  {author} {\bibinfo {author} {\bibfnamefont {F.~D.~M.}\
  \bibnamefont {Haldane}},\ }\bibfield  {title} {\bibinfo {title} {Model for a
  quantum {Hall} effect without {Landau} levels: Condensed-matter realization
  of the ``parity anomaly''},\ }\href
  {https://journals.aps.org/prl/abstract/10.1103/PhysRevLett.61.2015}
  {\bibfield  {journal} {\bibinfo  {journal} {Phys. Rev. Lett.}\ }\textbf
  {\bibinfo {volume} {61}},\ \bibinfo {pages} {2015} (\bibinfo {year}
  {1988})}\BibitemShut {NoStop}%
\bibitem [{\citenamefont {Liu}\ \emph {et~al.}(2008)\citenamefont {Liu},
  \citenamefont {Qi}, \citenamefont {Dai}, \citenamefont {Fang},\ and\
  \citenamefont {Zhang}}]{liu2008quantum}%
  \BibitemOpen
  \bibfield  {author} {\bibinfo {author} {\bibfnamefont {C.-X.}\ \bibnamefont
  {Liu}}, \bibinfo {author} {\bibfnamefont {X.-L.}\ \bibnamefont {Qi}},
  \bibinfo {author} {\bibfnamefont {X.}~\bibnamefont {Dai}}, \bibinfo {author}
  {\bibfnamefont {Z.}~\bibnamefont {Fang}},\ and\ \bibinfo {author}
  {\bibfnamefont {S.-C.}\ \bibnamefont {Zhang}},\ }\bibfield  {title} {\bibinfo
  {title} {Quantum anomalous {Hall} effect in $\mathrm{Hg_{1-y}Mn_yTe}$ quantum
  wells},\ }\href
  {https://journals.aps.org/prl/abstract/10.1103/PhysRevLett.101.146802}
  {\bibfield  {journal} {\bibinfo  {journal} {Phys. Rev. Lett.}\ }\textbf
  {\bibinfo {volume} {101}},\ \bibinfo {pages} {146802} (\bibinfo {year}
  {2008})}\BibitemShut {NoStop}%
\bibitem [{\citenamefont {Qi}\ \emph {et~al.}(2008)\citenamefont {Qi},
  \citenamefont {Hughes},\ and\ \citenamefont {Zhang}}]{qi2008topological}%
  \BibitemOpen
  \bibfield  {author} {\bibinfo {author} {\bibfnamefont {X.-L.}\ \bibnamefont
  {Qi}}, \bibinfo {author} {\bibfnamefont {T.~L.}\ \bibnamefont {Hughes}},\
  and\ \bibinfo {author} {\bibfnamefont {S.-C.}\ \bibnamefont {Zhang}},\
  }\bibfield  {title} {\bibinfo {title} {Topological field theory of
  time-reversal invariant insulators},\ }\href
  {https://journals.aps.org/prb/abstract/10.1103/PhysRevB.78.195424} {\bibfield
   {journal} {\bibinfo  {journal} {Phys. Rev. B}\ }\textbf {\bibinfo {volume}
  {78}},\ \bibinfo {pages} {195424} (\bibinfo {year} {2008})}\BibitemShut
  {NoStop}%
\bibitem [{\citenamefont {Yu}\ \emph {et~al.}(2010)\citenamefont {Yu},
  \citenamefont {Zhang}, \citenamefont {Zhang}, \citenamefont {Zhang},
  \citenamefont {Dai},\ and\ \citenamefont {Fang}}]{yu2010quantized}%
  \BibitemOpen
  \bibfield  {author} {\bibinfo {author} {\bibfnamefont {R.}~\bibnamefont
  {Yu}}, \bibinfo {author} {\bibfnamefont {W.}~\bibnamefont {Zhang}}, \bibinfo
  {author} {\bibfnamefont {H.-J.}\ \bibnamefont {Zhang}}, \bibinfo {author}
  {\bibfnamefont {S.-C.}\ \bibnamefont {Zhang}}, \bibinfo {author}
  {\bibfnamefont {X.}~\bibnamefont {Dai}},\ and\ \bibinfo {author}
  {\bibfnamefont {Z.}~\bibnamefont {Fang}},\ }\bibfield  {title} {\bibinfo
  {title} {Quantized anomalous {Hall} effect in magnetic topological
  insulators},\ }\href {https://science.sciencemag.org/content/329/5987/61}
  {\bibfield  {journal} {\bibinfo  {journal} {Science}\ }\textbf {\bibinfo
  {volume} {329}},\ \bibinfo {pages} {61} (\bibinfo {year} {2010})}\BibitemShut
  {NoStop}%
\bibitem [{\citenamefont {Chang}\ \emph {et~al.}(2013)\citenamefont {Chang},
  \citenamefont {Zhang}, \citenamefont {Feng}, \citenamefont {Shen},
  \citenamefont {Zhang}, \citenamefont {Guo}, \citenamefont {Li}, \citenamefont
  {Ou}, \citenamefont {Wei}, \citenamefont {Wang} \emph
  {et~al.}}]{chang2013experimental}%
  \BibitemOpen
  \bibfield  {author} {\bibinfo {author} {\bibfnamefont {C.-Z.}\ \bibnamefont
  {Chang}}, \bibinfo {author} {\bibfnamefont {J.}~\bibnamefont {Zhang}},
  \bibinfo {author} {\bibfnamefont {X.}~\bibnamefont {Feng}}, \bibinfo {author}
  {\bibfnamefont {J.}~\bibnamefont {Shen}}, \bibinfo {author} {\bibfnamefont
  {Z.}~\bibnamefont {Zhang}}, \bibinfo {author} {\bibfnamefont
  {M.}~\bibnamefont {Guo}}, \bibinfo {author} {\bibfnamefont {K.}~\bibnamefont
  {Li}}, \bibinfo {author} {\bibfnamefont {Y.}~\bibnamefont {Ou}}, \bibinfo
  {author} {\bibfnamefont {P.}~\bibnamefont {Wei}}, \bibinfo {author}
  {\bibfnamefont {L.-L.}\ \bibnamefont {Wang}}, \emph {et~al.},\ }\bibfield
  {title} {\bibinfo {title} {Experimental observation of the quantum anomalous
  {Hall} effect in a magnetic topological insulator},\ }\href
  {https://science.sciencemag.org/content/340/6129/167} {\bibfield  {journal}
  {\bibinfo  {journal} {Science}\ }\textbf {\bibinfo {volume} {340}},\ \bibinfo
  {pages} {167} (\bibinfo {year} {2013})}\BibitemShut {NoStop}%
\bibitem [{\citenamefont {Chang}\ \emph
  {et~al.}(2015{\natexlab{a}})\citenamefont {Chang}, \citenamefont {Zhao},
  \citenamefont {Kim}, \citenamefont {Zhang}, \citenamefont {Assaf},
  \citenamefont {Heiman}, \citenamefont {Zhang}, \citenamefont {Liu},
  \citenamefont {Chan},\ and\ \citenamefont {Moodera}}]{chang2015high}%
  \BibitemOpen
  \bibfield  {author} {\bibinfo {author} {\bibfnamefont {C.-Z.}\ \bibnamefont
  {Chang}}, \bibinfo {author} {\bibfnamefont {W.}~\bibnamefont {Zhao}},
  \bibinfo {author} {\bibfnamefont {D.~Y.}\ \bibnamefont {Kim}}, \bibinfo
  {author} {\bibfnamefont {H.}~\bibnamefont {Zhang}}, \bibinfo {author}
  {\bibfnamefont {B.~A.}\ \bibnamefont {Assaf}}, \bibinfo {author}
  {\bibfnamefont {D.}~\bibnamefont {Heiman}}, \bibinfo {author} {\bibfnamefont
  {S.-C.}\ \bibnamefont {Zhang}}, \bibinfo {author} {\bibfnamefont
  {C.}~\bibnamefont {Liu}}, \bibinfo {author} {\bibfnamefont {M.~H.}\
  \bibnamefont {Chan}},\ and\ \bibinfo {author} {\bibfnamefont {J.~S.}\
  \bibnamefont {Moodera}},\ }\bibfield  {title} {\bibinfo {title}
  {High-precision realization of robust quantum anomalous {Hall} state in a
  hard ferromagnetic topological insulator},\ }\href
  {https://www.nature.com/articles/nmat4204} {\bibfield  {journal} {\bibinfo
  {journal} {Nat. Mater.}\ }\textbf {\bibinfo {volume} {14}},\ \bibinfo {pages}
  {473} (\bibinfo {year} {2015}{\natexlab{a}})}\BibitemShut {NoStop}%
\bibitem [{\citenamefont {Checkelsky}\ \emph {et~al.}(2014)\citenamefont
  {Checkelsky}, \citenamefont {Yoshimi}, \citenamefont {Tsukazaki},
  \citenamefont {Takahashi}, \citenamefont {Kozuka}, \citenamefont {Falson},
  \citenamefont {Kawasaki},\ and\ \citenamefont
  {Tokura}}]{checkelsky2014trajectory}%
  \BibitemOpen
  \bibfield  {author} {\bibinfo {author} {\bibfnamefont {J.}~\bibnamefont
  {Checkelsky}}, \bibinfo {author} {\bibfnamefont {R.}~\bibnamefont {Yoshimi}},
  \bibinfo {author} {\bibfnamefont {A.}~\bibnamefont {Tsukazaki}}, \bibinfo
  {author} {\bibfnamefont {K.}~\bibnamefont {Takahashi}}, \bibinfo {author}
  {\bibfnamefont {Y.}~\bibnamefont {Kozuka}}, \bibinfo {author} {\bibfnamefont
  {J.}~\bibnamefont {Falson}}, \bibinfo {author} {\bibfnamefont
  {M.}~\bibnamefont {Kawasaki}},\ and\ \bibinfo {author} {\bibfnamefont
  {Y.}~\bibnamefont {Tokura}},\ }\bibfield  {title} {\bibinfo {title}
  {Trajectory of the anomalous {Hall} effect towards the quantized state in a
  ferromagnetic topological insulator},\ }\href
  {https://www.nature.com/articles/nphys3053} {\bibfield  {journal} {\bibinfo
  {journal} {Nat. Phys.}\ }\textbf {\bibinfo {volume} {10}},\ \bibinfo {pages}
  {731} (\bibinfo {year} {2014})}\BibitemShut {NoStop}%
\bibitem [{\citenamefont {Kou}\ \emph {et~al.}(2014)\citenamefont {Kou},
  \citenamefont {Guo}, \citenamefont {Fan}, \citenamefont {Pan}, \citenamefont
  {Lang}, \citenamefont {Jiang}, \citenamefont {Shao}, \citenamefont {Nie},
  \citenamefont {Murata}, \citenamefont {Tang} \emph {et~al.}}]{kou2014scale}%
  \BibitemOpen
  \bibfield  {author} {\bibinfo {author} {\bibfnamefont {X.}~\bibnamefont
  {Kou}}, \bibinfo {author} {\bibfnamefont {S.-T.}\ \bibnamefont {Guo}},
  \bibinfo {author} {\bibfnamefont {Y.}~\bibnamefont {Fan}}, \bibinfo {author}
  {\bibfnamefont {L.}~\bibnamefont {Pan}}, \bibinfo {author} {\bibfnamefont
  {M.}~\bibnamefont {Lang}}, \bibinfo {author} {\bibfnamefont {Y.}~\bibnamefont
  {Jiang}}, \bibinfo {author} {\bibfnamefont {Q.}~\bibnamefont {Shao}},
  \bibinfo {author} {\bibfnamefont {T.}~\bibnamefont {Nie}}, \bibinfo {author}
  {\bibfnamefont {K.}~\bibnamefont {Murata}}, \bibinfo {author} {\bibfnamefont
  {J.}~\bibnamefont {Tang}}, \emph {et~al.},\ }\bibfield  {title} {\bibinfo
  {title} {Scale-invariant quantum anomalous {Hall} effect in magnetic
  topological insulators beyond the two-dimensional limit},\ }\href
  {https://journals.aps.org/prl/abstract/10.1103/PhysRevLett.113.137201}
  {\bibfield  {journal} {\bibinfo  {journal} {Phys. Rev. Lett.}\ }\textbf
  {\bibinfo {volume} {113}},\ \bibinfo {pages} {137201} (\bibinfo {year}
  {2014})}\BibitemShut {NoStop}%
\bibitem [{\citenamefont {Mogi}\ \emph {et~al.}(2015)\citenamefont {Mogi},
  \citenamefont {Yoshimi}, \citenamefont {Tsukazaki}, \citenamefont {Yasuda},
  \citenamefont {Kozuka}, \citenamefont {Takahashi}, \citenamefont {Kawasaki},\
  and\ \citenamefont {Tokura}}]{mogi2015magnetic}%
  \BibitemOpen
  \bibfield  {author} {\bibinfo {author} {\bibfnamefont {M.}~\bibnamefont
  {Mogi}}, \bibinfo {author} {\bibfnamefont {R.}~\bibnamefont {Yoshimi}},
  \bibinfo {author} {\bibfnamefont {A.}~\bibnamefont {Tsukazaki}}, \bibinfo
  {author} {\bibfnamefont {K.}~\bibnamefont {Yasuda}}, \bibinfo {author}
  {\bibfnamefont {Y.}~\bibnamefont {Kozuka}}, \bibinfo {author} {\bibfnamefont
  {K.}~\bibnamefont {Takahashi}}, \bibinfo {author} {\bibfnamefont
  {M.}~\bibnamefont {Kawasaki}},\ and\ \bibinfo {author} {\bibfnamefont
  {Y.}~\bibnamefont {Tokura}},\ }\bibfield  {title} {\bibinfo {title} {Magnetic
  modulation doping in topological insulators toward higher-temperature quantum
  anomalous {Hall} effect},\ }\href
  {https://aip.scitation.org/doi/10.1063/1.4935075} {\bibfield  {journal}
  {\bibinfo  {journal} {Appl. Phys. Lett.}\ }\textbf {\bibinfo {volume}
  {107}},\ \bibinfo {pages} {182401} (\bibinfo {year} {2015})}\BibitemShut
  {NoStop}%
\bibitem [{\citenamefont {Neubauer}\ \emph {et~al.}(2013)\citenamefont
  {Neubauer}, \citenamefont {Pfleiderer}, \citenamefont {Binz}, \citenamefont
  {Rosch}, \citenamefont {Ritz}, \citenamefont {Niklowitz},\ and\ \citenamefont
  {B{\"o}ni}}]{neubauer2013erratum}%
  \BibitemOpen
  \bibfield  {author} {\bibinfo {author} {\bibfnamefont {A.}~\bibnamefont
  {Neubauer}}, \bibinfo {author} {\bibfnamefont {C.}~\bibnamefont
  {Pfleiderer}}, \bibinfo {author} {\bibfnamefont {B.}~\bibnamefont {Binz}},
  \bibinfo {author} {\bibfnamefont {A.}~\bibnamefont {Rosch}}, \bibinfo
  {author} {\bibfnamefont {R.}~\bibnamefont {Ritz}}, \bibinfo {author}
  {\bibfnamefont {P. ~G.}~\bibnamefont {Niklowitz}},\ and\ \bibinfo {author}
  {\bibfnamefont {P.}~\bibnamefont {B{\"o}ni}},\ }\bibfield  {title} {\bibinfo
  {title} {Erratum: Topological {Hall} effect in the {A} phase of {MnSi}},\
  }\href {https://journals.aps.org/prl/pdf/10.1103/PhysRevLett.110.209902}
  {\bibfield  {journal} {\bibinfo  {journal} {Phys. Rev. Lett.}\ }\textbf
  {\bibinfo {volume} {110}},\ \bibinfo {pages} {209902(E)} (\bibinfo {year}
  {2013})}\BibitemShut {NoStop}%
\bibitem [{\citenamefont {Lee}\ \emph {et~al.}(2009)\citenamefont {Lee},
  \citenamefont {Kang}, \citenamefont {Onose}, \citenamefont {Tokura},\ and\
  \citenamefont {Ong}}]{lee2009unusual}%
  \BibitemOpen
  \bibfield  {author} {\bibinfo {author} {\bibfnamefont {M.}~\bibnamefont
  {Lee}}, \bibinfo {author} {\bibfnamefont {W.}~\bibnamefont {Kang}}, \bibinfo
  {author} {\bibfnamefont {Y.}~\bibnamefont {Onose}}, \bibinfo {author}
  {\bibfnamefont {Y.}~\bibnamefont {Tokura}},\ and\ \bibinfo {author}
  {\bibfnamefont {N.~P.}\ \bibnamefont {Ong}},\ }\bibfield  {title} {\bibinfo
  {title} {Unusual {Hall} effect anomaly in {MnSi} under pressure},\ }\href
  {https://journals.aps.org/prl/abstract/10.1103/PhysRevLett.102.186601}
  {\bibfield  {journal} {\bibinfo  {journal} {Phys. Rev. Lett.}\ }\textbf
  {\bibinfo {volume} {102}},\ \bibinfo {pages} {186601} (\bibinfo {year}
  {2009})}\BibitemShut {NoStop}%
\bibitem [{\citenamefont {Kanazawa}\ \emph {et~al.}(2011)\citenamefont
  {Kanazawa}, \citenamefont {Onose}, \citenamefont {Arima}, \citenamefont
  {Okuyama}, \citenamefont {Ohoyama}, \citenamefont {Wakimoto}, \citenamefont
  {Kakurai}, \citenamefont {Ishiwata},\ and\ \citenamefont
  {Tokura}}]{kanazawa2011large}%
  \BibitemOpen
  \bibfield  {author} {\bibinfo {author} {\bibfnamefont {N.}~\bibnamefont
  {Kanazawa}}, \bibinfo {author} {\bibfnamefont {Y.}~\bibnamefont {Onose}},
  \bibinfo {author} {\bibfnamefont {T.}~\bibnamefont {Arima}}, \bibinfo
  {author} {\bibfnamefont {D.}~\bibnamefont {Okuyama}}, \bibinfo {author}
  {\bibfnamefont {K.}~\bibnamefont {Ohoyama}}, \bibinfo {author} {\bibfnamefont
  {S.}~\bibnamefont {Wakimoto}}, \bibinfo {author} {\bibfnamefont
  {K.}~\bibnamefont {Kakurai}}, \bibinfo {author} {\bibfnamefont
  {S.}~\bibnamefont {Ishiwata}},\ and\ \bibinfo {author} {\bibfnamefont
  {Y.}~\bibnamefont {Tokura}},\ }\bibfield  {title} {\bibinfo {title} {Large
  topological {Hall} effect in a short-period helimagnet mnge},\ }\href
  {https://journals.aps.org/prl/abstract/10.1103/PhysRevLett.106.156603}
  {\bibfield  {journal} {\bibinfo  {journal} {Phys. Rev. Lett.}\ }\textbf
  {\bibinfo {volume} {106}},\ \bibinfo {pages} {156603} (\bibinfo {year}
  {2011})}\BibitemShut {NoStop}%
\bibitem [{\citenamefont {Huang}\ and\ \citenamefont
  {Chien}(2012)}]{huang2012extended}%
  \BibitemOpen
  \bibfield  {author} {\bibinfo {author} {\bibfnamefont {S.~X.}~\bibnamefont
  {Huang}}\ and\ \bibinfo {author} {\bibfnamefont {C. ~L.}~\bibnamefont {Chien}},\
  }\bibfield  {title} {\bibinfo {title} {Extended skyrmion phase in epitaxial
  {FeGe} (111) thin films},\ }\href
  {https://journals.aps.org/prl/abstract/10.1103/PhysRevLett.108.267201}
  {\bibfield  {journal} {\bibinfo  {journal} {Phys. Rev. Lett.}\ }\textbf
  {\bibinfo {volume} {108}},\ \bibinfo {pages} {267201} (\bibinfo {year}
  {2012})}\BibitemShut {NoStop}%
\bibitem [{\citenamefont {Matsuno}\ \emph {et~al.}(2016)\citenamefont
  {Matsuno}, \citenamefont {Ogawa}, \citenamefont {Yasuda}, \citenamefont
  {Kagawa}, \citenamefont {Koshibae}, \citenamefont {Nagaosa}, \citenamefont
  {Tokura},\ and\ \citenamefont {Kawasaki}}]{matsuno2016interface}%
  \BibitemOpen
  \bibfield  {author} {\bibinfo {author} {\bibfnamefont {J.}~\bibnamefont
  {Matsuno}}, \bibinfo {author} {\bibfnamefont {N.}~\bibnamefont {Ogawa}},
  \bibinfo {author} {\bibfnamefont {K.}~\bibnamefont {Yasuda}}, \bibinfo
  {author} {\bibfnamefont {F.}~\bibnamefont {Kagawa}}, \bibinfo {author}
  {\bibfnamefont {W.}~\bibnamefont {Koshibae}}, \bibinfo {author}
  {\bibfnamefont {N.}~\bibnamefont {Nagaosa}}, \bibinfo {author} {\bibfnamefont
  {Y.}~\bibnamefont {Tokura}},\ and\ \bibinfo {author} {\bibfnamefont
  {M.}~\bibnamefont {Kawasaki}},\ }\bibfield  {title} {\bibinfo {title}
  {Interface-driven topological hall effect in {SrRuO3-SrIrO3} bilayer},\
  }\href {https://advances.sciencemag.org/content/2/7/e1600304} {\bibfield
  {journal} {\bibinfo  {journal} {Sci. Adv.}\ }\textbf {\bibinfo {volume}
  {2}},\ \bibinfo {pages} {e1600304} (\bibinfo {year} {2016})}\BibitemShut
  {NoStop}%
\bibitem [{\citenamefont {Ohuchi}\ \emph {et~al.}(2018)\citenamefont {Ohuchi},
  \citenamefont {Matsuno}, \citenamefont {Ogawa}, \citenamefont {Kozuka},
  \citenamefont {Uchida}, \citenamefont {Tokura},\ and\ \citenamefont
  {Kawasaki}}]{ohuchi2018electric}%
  \BibitemOpen
  \bibfield  {author} {\bibinfo {author} {\bibfnamefont {Y.}~\bibnamefont
  {Ohuchi}}, \bibinfo {author} {\bibfnamefont {J.}~\bibnamefont {Matsuno}},
  \bibinfo {author} {\bibfnamefont {N.}~\bibnamefont {Ogawa}}, \bibinfo
  {author} {\bibfnamefont {Y.}~\bibnamefont {Kozuka}}, \bibinfo {author}
  {\bibfnamefont {M.}~\bibnamefont {Uchida}}, \bibinfo {author} {\bibfnamefont
  {Y.}~\bibnamefont {Tokura}},\ and\ \bibinfo {author} {\bibfnamefont
  {M.}~\bibnamefont {Kawasaki}},\ }\bibfield  {title} {\bibinfo {title}
  {Electric-field control of anomalous and topological {Hall} effects in oxide
  bilayer thin films},\ }\href
  {https://www.nature.com/articles/s41467-017-02629-3} {\bibfield  {journal}
  {\bibinfo  {journal} {Nat. Commun.}\ }\textbf {\bibinfo {volume} {9}},\
  \bibinfo {pages} {213} (\bibinfo {year} {2018})}\BibitemShut {NoStop}%
\bibitem [{\citenamefont {Yasuda}\ \emph {et~al.}(2016)\citenamefont {Yasuda},
  \citenamefont {Wakatsuki}, \citenamefont {Morimoto}, \citenamefont {Yoshimi},
  \citenamefont {Tsukazaki}, \citenamefont {Takahashi}, \citenamefont {Ezawa},
  \citenamefont {Kawasaki}, \citenamefont {Nagaosa},\ and\ \citenamefont
  {Tokura}}]{yasuda2016geometric}%
  \BibitemOpen
  \bibfield  {author} {\bibinfo {author} {\bibfnamefont {K.}~\bibnamefont
  {Yasuda}}, \bibinfo {author} {\bibfnamefont {R.}~\bibnamefont {Wakatsuki}},
  \bibinfo {author} {\bibfnamefont {T.}~\bibnamefont {Morimoto}}, \bibinfo
  {author} {\bibfnamefont {R.}~\bibnamefont {Yoshimi}}, \bibinfo {author}
  {\bibfnamefont {A.}~\bibnamefont {Tsukazaki}}, \bibinfo {author}
  {\bibfnamefont {K.}~\bibnamefont {Takahashi}}, \bibinfo {author}
  {\bibfnamefont {M.}~\bibnamefont {Ezawa}}, \bibinfo {author} {\bibfnamefont
  {M.}~\bibnamefont {Kawasaki}}, \bibinfo {author} {\bibfnamefont
  {N.}~\bibnamefont {Nagaosa}},\ and\ \bibinfo {author} {\bibfnamefont
  {Y.}~\bibnamefont {Tokura}},\ }\bibfield  {title} {\bibinfo {title}
  {Geometric {Hall} effects in topological insulator heterostructures},\ }\href
  {https://www.nature.com/articles/nphys3671} {\bibfield  {journal} {\bibinfo
  {journal} {Nat. Phys.}\ }\textbf {\bibinfo {volume} {12}},\ \bibinfo {pages}
  {555} (\bibinfo {year} {2016})}\BibitemShut {NoStop}%
\bibitem [{\citenamefont {Liu}\ \emph {et~al.}(2017)\citenamefont {Liu},
  \citenamefont {Zang}, \citenamefont {Ruan}, \citenamefont {Gong},
  \citenamefont {He}, \citenamefont {Ma}, \citenamefont {Xue},\ and\
  \citenamefont {Wang}}]{liu2017dimensional}%
  \BibitemOpen
  \bibfield  {author} {\bibinfo {author} {\bibfnamefont {C.}~\bibnamefont
  {Liu}}, \bibinfo {author} {\bibfnamefont {Y.}~\bibnamefont {Zang}}, \bibinfo
  {author} {\bibfnamefont {W.}~\bibnamefont {Ruan}}, \bibinfo {author}
  {\bibfnamefont {Y.}~\bibnamefont {Gong}}, \bibinfo {author} {\bibfnamefont
  {K.}~\bibnamefont {He}}, \bibinfo {author} {\bibfnamefont {X.}~\bibnamefont
  {Ma}}, \bibinfo {author} {\bibfnamefont {Q.-K.}\ \bibnamefont {Xue}},\ and\
  \bibinfo {author} {\bibfnamefont {Y.}~\bibnamefont {Wang}},\ }\bibfield
  {title} {\bibinfo {title} {Dimensional crossover-induced topological {Hall}
  effect in a magnetic topological insulator},\ }\href
  {https://journals.aps.org/prl/abstract/10.1103/PhysRevLett.119.176809}
  {\bibfield  {journal} {\bibinfo  {journal} {Phys. Rev. Lett.}\ }\textbf
  {\bibinfo {volume} {119}},\ \bibinfo {pages} {176809} (\bibinfo {year}
  {2017})}\BibitemShut {NoStop}%
\bibitem [{\citenamefont {Wang}\ \emph {et~al.}(2021)\citenamefont {Wang},
  \citenamefont {Zhao}, \citenamefont {Wang}, \citenamefont {Daniels},
  \citenamefont {Chang}, \citenamefont {Zang}, \citenamefont {Xiao},\ and\
  \citenamefont {Wu}}]{doi:10.1021/Wang_nanolett2021}%
  \BibitemOpen
  \bibfield  {author} {\bibinfo {author} {\bibfnamefont {W.}~\bibnamefont
  {Wang}}, \bibinfo {author} {\bibfnamefont {Y.-F.}\ \bibnamefont {Zhao}},
  \bibinfo {author} {\bibfnamefont {F.}~\bibnamefont {Wang}}, \bibinfo {author}
  {\bibfnamefont {M.~W.}\ \bibnamefont {Daniels}}, \bibinfo {author}
  {\bibfnamefont {C.-Z.}\ \bibnamefont {Chang}}, \bibinfo {author}
  {\bibfnamefont {J.}~\bibnamefont {Zang}}, \bibinfo {author} {\bibfnamefont
  {D.}~\bibnamefont {Xiao}},\ and\ \bibinfo {author} {\bibfnamefont
  {W.}~\bibnamefont {Wu}},\ }\bibfield  {title} {\bibinfo {title}
  {Chiral-bubble-induced topological {Hall} effect in ferromagnetic topological
  insulator heterostructures},\ }\href
  {https://doi.org/10.1021/acs.nanolett.0c04567} {\bibfield  {journal}
  {\bibinfo  {journal} {Nano Lett.}\ }\textbf {\bibinfo {volume} {21}},\
  \bibinfo {pages} {1108} (\bibinfo {year} {2021})}\BibitemShut {NoStop}%
\bibitem [{\citenamefont {Li}\ \emph {et~al.}(2021)\citenamefont {Li},
  \citenamefont {Ding}, \citenamefont {Zhang}, \citenamefont {Kally},
  \citenamefont {Pillsbury}, \citenamefont {Heinonen}, \citenamefont {Rimal},
  \citenamefont {Bi}, \citenamefont {DeMann}, \citenamefont {Field},
  \citenamefont {Wang}, \citenamefont {Tang}, \citenamefont {Jiang},
  \citenamefont {Hoffmann}, \citenamefont {Samarth},\ and\ \citenamefont
  {Wu}}]{Li_nanolett2021}%
  \BibitemOpen
  \bibfield  {author} {\bibinfo {author} {\bibfnamefont {P.}~\bibnamefont
  {Li}}, \bibinfo {author} {\bibfnamefont {J.}~\bibnamefont {Ding}}, \bibinfo
  {author} {\bibfnamefont {S.~S.-L.}\ \bibnamefont {Zhang}}, \bibinfo {author}
  {\bibfnamefont {J.}~\bibnamefont {Kally}}, \bibinfo {author} {\bibfnamefont
  {T.}~\bibnamefont {Pillsbury}}, \bibinfo {author} {\bibfnamefont {O.~G.}\
  \bibnamefont {Heinonen}}, \bibinfo {author} {\bibfnamefont {G.}~\bibnamefont
  {Rimal}}, \bibinfo {author} {\bibfnamefont {C.}~\bibnamefont {Bi}}, \bibinfo
  {author} {\bibfnamefont {A.}~\bibnamefont {DeMann}}, \bibinfo {author}
  {\bibfnamefont {S.~B.}\ \bibnamefont {Field}}, \bibinfo {author}
  {\bibfnamefont {W.}~\bibnamefont {Wang}}, \bibinfo {author} {\bibfnamefont
  {J.}~\bibnamefont {Tang}}, \bibinfo {author} {\bibfnamefont {J.~S.}\
  \bibnamefont {Jiang}}, \bibinfo {author} {\bibfnamefont {A.}~\bibnamefont
  {Hoffmann}}, \bibinfo {author} {\bibfnamefont {N.}~\bibnamefont {Samarth}},\
  and\ \bibinfo {author} {\bibfnamefont {M.}~\bibnamefont {Wu}},\ }\bibfield
  {title} {\bibinfo {title} {Topological {Hall} effect in a topological
  insulator interfaced with a magnetic insulator},\ }\href
  {https://doi.org/10.1021/acs.nanolett.0c03195} {\bibfield  {journal}
  {\bibinfo  {journal} {Nano Lett.}\ }\textbf {\bibinfo {volume} {21}},\
  \bibinfo {pages} {84} (\bibinfo {year} {2021})}\BibitemShut {NoStop}%
\bibitem [{\citenamefont {Jiang}\ \emph {et~al.}(2020)\citenamefont {Jiang},
  \citenamefont {Xiao}, \citenamefont {Wang}, \citenamefont {Shin},
  \citenamefont {Andreoli}, \citenamefont {Zhang}, \citenamefont {Xiao},
  \citenamefont {Zhao}, \citenamefont {Kayyalha}, \citenamefont {Zhang},
  \citenamefont {Wang}, \citenamefont {Zang}, \citenamefont {Liu},
  \citenamefont {Samarth}, \citenamefont {Chan},\ and\ \citenamefont
  {Chang}}]{jiang2020concurrence}%
  \BibitemOpen
  \bibfield  {author} {\bibinfo {author} {\bibfnamefont {J.}~\bibnamefont
  {Jiang}}, \bibinfo {author} {\bibfnamefont {D.}~\bibnamefont {Xiao}},
  \bibinfo {author} {\bibfnamefont {F.}~\bibnamefont {Wang}}, \bibinfo {author}
  {\bibfnamefont {J.-H.}\ \bibnamefont {Shin}}, \bibinfo {author}
  {\bibfnamefont {D.}~\bibnamefont {Andreoli}}, \bibinfo {author}
  {\bibfnamefont {J.}~\bibnamefont {Zhang}}, \bibinfo {author} {\bibfnamefont
  {R.}~\bibnamefont {Xiao}}, \bibinfo {author} {\bibfnamefont {Y.-F.}\
  \bibnamefont {Zhao}}, \bibinfo {author} {\bibfnamefont {M.}~\bibnamefont
  {Kayyalha}}, \bibinfo {author} {\bibfnamefont {L.}~\bibnamefont {Zhang}},
  \bibinfo {author} {\bibfnamefont {K.}~\bibnamefont {Wang}}, \bibinfo {author}
  {\bibfnamefont {J.}~\bibnamefont {Zang}}, \bibinfo {author} {\bibfnamefont
  {C.}~\bibnamefont {Liu}}, \bibinfo {author} {\bibfnamefont {N.}~\bibnamefont
  {Samarth}}, \bibinfo {author} {\bibfnamefont {M.~H.~W.}\ \bibnamefont
  {Chan}},\ and\ \bibinfo {author} {\bibfnamefont {C.-Z.}\ \bibnamefont
  {Chang}},\ }\bibfield  {title} {\bibinfo {title} {Concurrence of quantum
  anomalous {Hall} and topological {Hall} effects in magnetic topological
  insulator sandwich heterostructures},\ }\href
  {https://www.nature.com/articles/s41563-020-0605-z} {\bibfield  {journal}
  {\bibinfo  {journal} {Nat. Mat.}\ }\textbf {\bibinfo {volume} {19}},\
  \bibinfo {pages} {732} (\bibinfo {year} {2020})}\BibitemShut {NoStop}%
\bibitem [{\citenamefont {Chang}\ \emph
  {et~al.}(2015{\natexlab{b}})\citenamefont {Chang}, \citenamefont {Zhao},
  \citenamefont {Kim}, \citenamefont {Wei},
  \citenamefont {Jain}, , \citenamefont {Liu}, \citenamefont {Chan},\ and\
  \citenamefont {Moodera}}]{chang2015PRL}%
  \BibitemOpen
  \bibfield  {author} {\bibinfo {author} {\bibfnamefont {C.-Z.}\ \bibnamefont
  {Chang}}, \bibinfo {author} {\bibfnamefont {W.}~\bibnamefont {Zhao}},
  \bibinfo {author} {\bibfnamefont {D.~Y.}\ \bibnamefont {Kim}}, \bibinfo
  {author} {\bibfnamefont {P.}~\bibnamefont {Wei}}, \bibinfo {author} {\bibfnamefont
  {J.~K.}\ \bibnamefont {Jain}}, , \bibinfo {author} {\bibfnamefont
  {C.}~\bibnamefont {Liu}}, \bibinfo {author} {\bibfnamefont {M.~H.~W.}\
  \bibnamefont {Chan}},\ and\ \bibinfo {author} {\bibfnamefont {J.~S.}\
  \bibnamefont {Moodera}},\ }\bibfield  {title} {\bibinfo {title} {Zero-field
  dissipationless chiral edge transport and the nature of dissipation in the
  quantum anomalous {Hall} state},\ }\href
  {https://journals.aps.org/prl/abstract/10.1103/PhysRevLett.115.057206}
  {\bibfield  {journal} {\bibinfo  {journal} {Phys. Rev. Lett.}\ }\textbf
  {\bibinfo {volume} {115}},\ \bibinfo {pages} {057206} (\bibinfo {year}
  {2015}{\natexlab{b}})}\BibitemShut {NoStop}%
\bibitem [{\citenamefont {Wang}\ \emph {et~al.}(2018)\citenamefont {Wang},
  \citenamefont {Ou}, \citenamefont {Liu}, \citenamefont {Wang}, \citenamefont
  {He}, \citenamefont {Xue},\ and\ \citenamefont {Wu}}]{Wang2018NP}%
  \BibitemOpen
  \bibfield  {author} {\bibinfo {author} {\bibfnamefont {W.}~\bibnamefont
  {Wang}}, \bibinfo {author} {\bibfnamefont {Y.}~\bibnamefont {Ou}}, \bibinfo
  {author} {\bibfnamefont {C.}~\bibnamefont {Liu}}, \bibinfo {author}
  {\bibfnamefont {Y.}~\bibnamefont {Wang}}, \bibinfo {author} {\bibfnamefont
  {K.}~\bibnamefont {He}}, \bibinfo {author} {\bibfnamefont {Q.-K.}\
  \bibnamefont {Xue}},\ and\ \bibinfo {author} {\bibfnamefont {W.}~\bibnamefont
  {Wu}},\ }\bibfield  {title} {\bibinfo {title} {Direct evidence of
  ferromagnetism in a quantum anomalous {Hall} system},\ }\href
  {https://www.nature.com/articles/s41567-018-0149-1} {\bibfield  {journal}
  {\bibinfo  {journal} {Nat. Phys.}\ }\textbf {\bibinfo {volume} {14}},\
  \bibinfo {pages} {791} (\bibinfo {year} {2018})}\BibitemShut {NoStop}%
\bibitem [{\citenamefont {Fijalkowski}\ \emph {et~al.}(2020)\citenamefont
  {Fijalkowski}, \citenamefont {Hartl}, \citenamefont {Winnerlein},
  \citenamefont {Mandal}, \citenamefont {Schreyeck}, \citenamefont {Brunner},
  \citenamefont {Gould},\ and\ \citenamefont
  {Molenkamp}}]{Filakowski_PhysRevX.10.011012}%
  \BibitemOpen
  \bibfield  {author} {\bibinfo {author} {\bibfnamefont {K.~M.}\ \bibnamefont
  {Fijalkowski}}, \bibinfo {author} {\bibfnamefont {M.}~\bibnamefont {Hartl}},
  \bibinfo {author} {\bibfnamefont {M.}~\bibnamefont {Winnerlein}}, \bibinfo
  {author} {\bibfnamefont {P.}~\bibnamefont {Mandal}}, \bibinfo {author}
  {\bibfnamefont {S.}~\bibnamefont {Schreyeck}}, \bibinfo {author}
  {\bibfnamefont {K.}~\bibnamefont {Brunner}}, \bibinfo {author} {\bibfnamefont
  {C.}~\bibnamefont {Gould}},\ and\ \bibinfo {author} {\bibfnamefont {L.~W.}\
  \bibnamefont {Molenkamp}},\ }\bibfield  {title} {\bibinfo {title}
  {Coexistence of surface and bulk ferromagnetism mimics skyrmion {Hall} effect
  in a topological insulator},\ }\href
  {https://doi.org/10.1103/PhysRevX.10.011012} {\bibfield  {journal} {\bibinfo
  {journal} {Phys. Rev. X}\ }\textbf {\bibinfo {volume} {10}},\ \bibinfo
  {pages} {011012} (\bibinfo {year} {2020})}\BibitemShut {NoStop}%
\end{thebibliography}

%apsrev4-2.bst 2019-01-14 (MD) hand-edited version of apsrev4-1.bst
%Control: key (0)
%Control: author (8) initials jnrlst
%Control: editor formatted (1) identically to author
%Control: production of article title (0) allowed
%Control: page (0) single
%Control: year (1) truncated
%Control: production of eprint (0) enabled
\providecommand{\noopsort}[1]{}\providecommand{\singleletter}[1]{#1}%

\end{document}